\documentclass[conference]{IEEEtran}
\IEEEoverridecommandlockouts
\usepackage{cite}
\usepackage{amsmath,amssymb,amsfonts}
\usepackage{algorithmic}
\usepackage{graphicx}
\usepackage{textcomp}
\def\BibTeX{{\rm B\kern-.05em{\sc i\kern-.025em b}\kern-.08em
		T\kern-.1667em\lower.7ex\hbox{E}\kern-.125emX}}

\usepackage[T1]{fontenc}
\usepackage{textcomp}
\usepackage[scaled=.92]{helvet} 
\usepackage{graphicx} 
\usepackage{balance}  
\usepackage{booktabs} 
\usepackage{ccicons}  
\usepackage{ragged2e} 
\usepackage[dvipsnames]{xcolor}
\usepackage{breakcites}
\usepackage{hyperref}

\makeatletter
\newcommand*\titleheader[1]{\gdef\@titleheader{#1}}
\AtBeginDocument{%
	\let\st@red@title\@title
	\def\@title{%
		\bgroup\normalfont\large\centering\@titleheader\par\egroup
		\vskip0.8em\st@red@title}
}
\makeatother

\title{Reclaiming Data: Overcoming App Identification Barriers for Exercising Data Protection Rights} 

\titleheader{\begin{footnotesize}Author preprint (accepted 20-Aug-18) To appear in the proceedings of the 4th Workshop on Legal and Technical Issues in \\Cloud and Pervasive Computing (IoT) [CLaw-18], \emph{UbiComp/ISWC'18 Adjunct,} \url{https://doi.org/10.1145/3267305.3274153}\end{footnotesize}}

\begin{document}

\def\plaintitle{Reclaiming Data: Overcoming App Identification Barriers for Exercising Data Protection Rights} 
\def\plainauthor{Chris Norval, Jennifer Cobbe, Heleen Janssen, Jatinder Singh}
\def\plainkeywords{identity management; data protection; privacy; GDPR; subject access rights; mobile applications}
\def\plaingeneralterms{Documentation, Standardization}

\author{
\IEEEauthorblockN{Chris Norval, Jennifer Cobbe, Heleen Janssen, Jatinder Singh}
\IEEEauthorblockA{\textit{Department of Computer Science \& Technology} \\
\textit{University of Cambridge, UK}\\
firstname.lastname@cl.cam.ac.uk}
}

\definecolor{linkColor}{RGB}{6,125,233}
\hypersetup{%
  pdftitle={\plaintitle},
  pdfauthor={\plainauthor},
  pdfkeywords={\plainkeywords},
  bookmarksnumbered,
  pdfstartview={FitH},
  colorlinks,
  citecolor=black,
  filecolor=black,
  linkcolor=black,
  urlcolor=linkColor,
  breaklinks=true,
  hidelinks=true
}

\newcommand{\notecn}[1]
	{{\color{blue}[{\bf Chris:} #1]}}
\newcommand{\notejs}[1]
	{{\color{brown}[{\bf Jat:} #1]}}
\newcommand{\notehj}[1]
	{{\color{violet}[{\bf Heleen:} #1]}}
\newcommand{\notejc}[1]
	{{\color{green}[{\bf Jen:} #1]}}
\newcommand{\change}[1]
	{{\color{orange}#1}}
\newcommand{\todo}[1]{{\color{purple}[{\bf To Do:} #1]}}
\newcommand{\priority}[1]{{\color{red}[{\bf Priority:} #1]}}
	

\maketitle


\begin{abstract}

Data protection regulations generally afford individuals certain rights over their personal data, including the rights to access, rectify, and delete the data held on them.
Exercising such rights naturally requires those with data management obligations (service providers) to be able to match an individual with their data. 
However, many mobile apps collect personal data, without requiring user registration or collecting details of a user's identity (email address, names, phone number, and so forth).
As a result, a user's ability to exercise their rights will be hindered without means for an individual to link themselves with this `nameless' data.
Current approaches often involve those seeking to exercise their legal rights having to give the app's provider more personal information, or even to  register for a service; both of which seem contrary to the spirit of data protection law.
This paper explores these concerns, and indicates simple means for facilitating data subject rights through both application and mobile platform (OS) design.

\end{abstract}




\section{Introduction}
A key aim of data protection law is to give individuals increased transparency, accountability, and control over their personal data. 
The EU's \emph{General Data Protection Regulation} (GDPR)~\cite{gdpr2016}\footnote{In this paper we focus on the GDPR, recognising that similar principles will be relevant to similar legal frameworks in other jurisdictions.} modernises and strengthens the rights of \emph{data subjects}\footnote{An identifiable natural person is one who can be identified, directly or indirectly, in particular by reference to an identifier such as a name, an identification number, location data, an online identifier or to one or more factors specific to the physical, physiological, genetic, mental, economic, cultural or social identity of that natural person. GDPR, Art 4(1).} in relation to how their personal data is collected and processed by \emph{data controllers}\footnote{The natural or legal person, public authority, agency or other body which, alone or jointly with others, determines the purposes and means of the processing of personal data. GDPR, Art 4(7).} -- regardless of which country that data controller is based.
The GDPR provides several rights which a data subject may wish to exercise. 
For example, the `right of access'\footnote{GDPR, Art 15.} gives the data subject the ability to access personal data held about themselves by a data controller.
Data subjects can also request the correction of incorrect data through the `right to rectification',\footnote{GDPR, Art 16.} or the deletion of their data through the `right to erasure' (commonly referred to as the `right to be forgotten'),\footnote{GDPR, Art 17.} among others.

Many mobile applications (\emph{`apps'}) entail the collection and (server-side) processing of data, which is often personal in nature and therefore subject to such rights. 
However, many of these apps operate without requiring registration (sign-in) mechanisms, and do not collect other identity-related information (email address, name, etc).
This situation leads to the challenge for individuals attempting to exercise their rights: \emph{how does a user (data subject) supply sufficient information to allow the app's provider to identify the user's data so as to action the request}?\footnote{By `app provider', we refer to the entity responsible for the provision of the service.} \footnote{By `identify', we refer to the act of technically matching a given user/device with their data through common identity information --- not the act of verifying the authenticity of a rights request once received.}

App providers will generally use some identifier(s) to uniquely differentiate data from a particular device\slash user -- typically those provided by the mobile platforms (operating systems). 
Yet, these identifiers are generally not accessible to end-users.
Without identifiers which are both available to end-users and app providers, it may be difficult (perhaps impossible) to link a user with their data on the app provider's systems.

Towards this, current approaches seem to involve the user having to supply additional data about themselves in order to exercise their rights. 
This might entail the user having to sign-up for the service in question or sign in to a third-party identity provider (e.g. Facebook) in order to bridge the divide.
Neither of these may be  desirable, not least because the user may be trying to limit the data held by the controller rather than provide them with more.
Indeed, such approaches 
appear contrary to the spirit of the data protection law.

This paper explores issues regarding user identification and exercising legal rights in the context of apps, suggesting simple ways forward for both app providers and mobile platforms. 
In line with other work (e.g.~\cite{veale2018}), we generally seek to raise awareness of how technology design impacts rights, highlighting the need to account for data subject rights as part of app development.

\section{Mobile Apps, Data Collection \& Subject Rights} \label{datacollection}

Many apps will involve the generation or collection of personal data which is sent back to the app's provider for processing.\footnote{
Though not all apps do this, it is  very common for apps to involve data transfer and processing on remote servers (cloud), as opposed keeping and processing all data only on the device.}
Yet, many users may be unaware of the nature and scale of such data collection, or might have general questions about how their data is being used.\footnote{Ofcom recently found that a majority of UK-based Internet users that responded to a survey were unaware that some apps \textit{``collect data on users' locations, or on what products or services interest them"}~\cite{ofcom2018}.}
As such, the rights afforded by the GDPR are powerful tools for transparency, accountability, and user agency.

While many of the GDPR's requirements are not new,\footnote{The GDPR's predecessor, the Data Protection Directive~\cite{dpd1995}, was passed in 1995 and contained most of the same data subject rights.} its renewed emphasis on data subject rights and the regulatory fines it provides for non-compliance have, for many, brought these issues into sharp focus. 
However, the GDPR's complexity, and the fact that many individuals and organisations are only now properly beginning to consider these issues, could mean that app providers are subject to the requirements of the GDPR without realising it.
For example, personal data is any data which relates to an identified or \emph{identifiable} natural person; that is, one who can be identified directly or indirectly.\footnote{GDPR, Art 4(1).} 
As such, data that does not directly include identity information about the individual can also fall within the scope of the GDPR~\cite{ico2018:anonymisation, koops2014}.\footnote{Research has repeatedly shown that even anonymized data can often be re-identified and attributed to specific individuals~\cite{tene2011}, and there is ongoing discussion about the extent to which non-personal data should even exist as a concept given that \textit{``in increasingly `smart' environments any information is likely to relate to a person in purpose or effect"}~\cite{purtova2018}.
}


In short, where the data can be linked to a particular individual --- whether they are known or identified to the app service provider or not --- it will be personal data against which GDPR rights can be exercised.
Yet, some app providers may not even realise that such data is within the scope of data protection regulations, or even what their data protection obligations are.
Consequently, the requirements of the GDPR may not have been considered during the design and development of such an app.
In turn, this may result in circumstances where the data subject has few means for identifying themselves so as to exercise their rights over that data.

\subsection{Qualifications and Exceptions}
Data controllers are generally obliged by the GDPR to facilitate the exercise of data subject rights, including taking the necessary technical and organisational measures to permit their exercise, and should not obstruct data subjects who are doing so.\footnote{GDPR, Arts 12 and 25; Recitals 59 and 78.} 
However, there are two circumstances in which this is not necessarily the case.
In each, the onus is on the data controller to demonstrate that the circumstance exists.

The first of these circumstances is in relation to services which do not require the identification in some way of individual end-users (i.e. where they are not providing a personalised service or linking data coming to and from a device as belonging to a particular app instance, etc.) and where the data controller can demonstrate that they aren't in a position to identify the data subject.\footnote{GDPR, Art 11.} This could be, for example, a weather app which includes coarse-grained location information in a request for forecast data and where the app provider only stores this information in aggregate.
In these cases, data controllers are under no obligation to provide means for data subjects to be identified, and certain data subject rights don't apply unless the data subject provides sufficient identifying information (once a data subject has identified themselves by providing the necessary information, the controller must according to Article 11(2) of the GDPR proceed to meet its request).\footnote{GDPR, Recital 57.}

Yet a distinction should be drawn between cases where a) the service provided does not need to collect identifying information (e.g. as above) and b) situations where the service \emph{does} need to identify users (say, to provide a personalised experience), but the data controller simply does not collect the information which would allow them to link users with their personal data. 
In the latter case, users would not be able to effectively exercise their data subject rights, and failure to fulfil data subject requests would be a breach of the GDPR.
Appropriate technical and organisational measures should therefore be implemented by the controller in order to meet the requirements of the GDPR.\footnote{GDPR, Art 25.}


The second circumstance in which data controllers may not be obliged to facilitate the exercise of data subject rights is where the data subject's request is manifestly unfounded or excessive (in these cases the data controller may refuse the request or may elect to charge a fee).\footnote{GDPR, Art 12(5).} For this to apply, as in relation to the first circumstance discussed above, the data controller must themselves demonstrate that the request is either manifestly unfounded or excessive; however, this may prove challenging, given the technical ease of linking data via queries for instance.

To summarise, \emph{there are many situations where data collected by apps --- even those which do not collect directly identifiable data --- are subject to the rights afforded by the GDPR}. 
Yet, as discussed in the Introduction, this is possible only where some shared identifier(s) (or some other means) exist to link a data subject with their data. 
Therefore, the technical attributes and capabilities of these identifiers are an important part of this discussion.

\section{Identifiers and devices} \label{identifiers}
Apps that involve `off-device' (i.e. server\slash cloud) data processing will typically use identifiers, of some form, against which a data `profile' is built. 
We use \emph{`profile'} to describe data associated with a particular user as seen through the eyes of the `off-device' service.
In practice, this could be a dataset associated with an identified, but otherwise unnamed, user of the app. 
Note that our usage is distinct from `profiling' as outlined in the GDPR, which defines it as \textit{``any form of automated processing of personal data consisting of the use of personal data to evaluate certain personal aspects relating to a natural person''}.\footnote{GDPR, Art 4(4).}

As discussed, where apps collect data that can directly identify an individual user (e.g. email address or username), it should generally be possible for the provider (when given identifying information by the data subject) to uncover the data\slash profile associated with that user. 
This might be, for instance, by way of simple database queries. 
This means that a data subject seeking to exercise their rights (should) need only to provide this identity information in order for the data controller to action their request (generally after some verification process).
Note that the internal identifier used by such apps to associate user data needn't necessarily be the identity information itself (e.g. the email address), as long as that identity information links to something with which that user's data is associated (e.g. an internal identifier string for the data profile).

For apps which do not require signing up or the provision of any identifiable data, generally some form of profile for a particular user will still be created.
This may be through an identifier which is specific to the device, such as a serial number, or it could be a programmatically generated identifier provided by the platform or the app's developer at the instance the app started for the first time.
Given that (a) many of these programmatic identifiers are neither comprehensible nor accessible to end-users, and (b) the app provider may have no other identity information, the key issue is the limited means with which individuals can assert their identity and thereby have app providers uncover which data profile belongs to them.
See Fig.~\ref{fig:identifiers} for an illustration.

\subsection{Platform managed identifiers}

App providers require identifiers to help manage the data associated with a user.
Though an app can create its own identifier,\footnote{For example, a random string generated on the first execution of an app that is used with future interactions with their server(s).} given the commonality and general need for such functionality, the mobile platforms have in-built programmatic methods to facilitate user\slash device identification.
In the earlier days of iOS and Android (<2013), many apps would use some form of alphanumerical string which was unique for the particular device, such as \texttt{Settings.Secure.ANDROID\_ID} (SSAID) on Android, or the \emph{Unique Device Identifier} (UDID) on iOS. 
These (generally) immutable device-centric identifiers were accessible to app developers, and every app on the device would see the same identifier value. 
Importantly, these identifiers were also accessible to the device owners, though at varying levels of ease.
In such situations, one could imagine a user providing their device identifier to an app provider to assist with their exercising of rights.

\begin{figure}[t!]
	\centering
	\includegraphics[width=.78\linewidth]{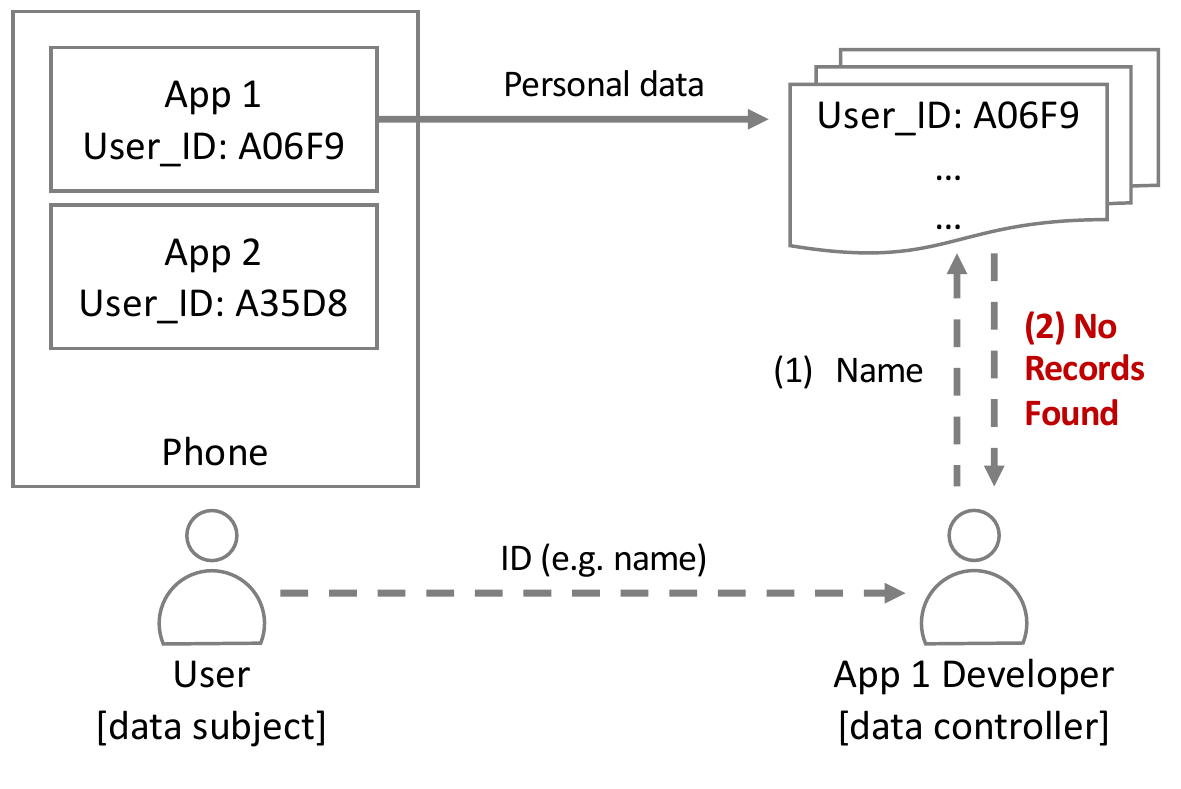}
	\caption{\small Example where a user generates and transmits personal data via an app to a data controller. 
		Since the user has no access to their User\_ID on App 1, and the app provider has no access to other identifiable data (e.g. name), it becomes difficult to match the user with their data to facilitate the user's data protection rights.}
	\label{fig:identifiers}
\end{figure}

The identifier landscape, however, has since evolved.
Recognising the privacy and security implications of allowing all apps to access the (same) identifiers that uniquely refer to a specific device, the iOS App Store review process began rejecting apps which accessed the UDID in 2012. This followed media reports that some app providers were sharing these identifiers with advertising companies for cross-app profiling purposes~\cite{lifehacker2012}.
Apple created a new identifier for iOS developers to use, \texttt{identifierForVendor} (IDFV), which is unique to each device \emph{and} developer account combination.\footnote{See: \texttt{identifierForVendor} on the Apple Developer Documentation~\cite{appledocs2018}.}
This allows a particular app provider to get the same identifier across \emph{their} applications but will give apps from different developer accounts a different value (e.g. see Fig.~\ref{fig:identifiers}), preventing the matching of a single device across different providers.

Like Apple, Google began discouraging the use of hardware identifiers on Android,\footnote{See: \textit{Best practices for unique identifiers} on the Android Developer Documentation~\cite{androiddocs2018}.} eventually changing how the SSAID identifier worked.
From Android O (v8, 2017), requesting the SSAID returned a different value for each app instance~\cite{androidblog2018},\footnote{Note the subtle differences between IDFV and (new) SSAID. IDFVs are unique for a given developer account and device combination, whereas SSAIDs are unique for a given app and user on a device.} and thus it could no longer be used for cross-app profiling purposes.

Yet it appears that neither the IDFV or the (Android O onwards) SSAID for a given app are accessible or visible to users, unless specifically implemented into the application itself.
And while the platforms do provide some identifiers which are end-user accessible, these typically come with restrictions.
For example, Apple and Google both provide resettable, user-accessible identifiers for advertising purposes: \texttt{advertisingIdentifier} (IDFA) on iOS,\footnote{See: \texttt{advertisingIdentifier} and \emph{ASIdentifierManager} on the Apple Developer Documentation~\cite{appledocs2018}.} and \emph{Advertising ID} on Android~\cite{androidpolicy2018}.
These allow app providers and advertisers to track users\slash devices for advertising and profiling purposes -- yet with added functionality that allows the user to control (i.e. reset or block the use of) their advertising identifier.

While these advertising identifiers may have been effective technical mechanisms for linking users with their data, \emph{both} iOS and Android discourage their use beyond advertising.
Google Play's (Android) Developer Policy Center forbids associating the Advertising ID with any personally identifiable information or persistent device identifiers~\cite{androidpolicy2018}.
The iOS App Store actively rejects apps which use the IDFA for purposes beyond advertising~\cite{techcrunch2014} (e.g. when an app requests this identifier but adverts aren't enabled within the app).
Apps which access these identifiers for uses other than advertising may find themselves rejected from the platform marketplace's review process.

The technical workings and permitted use of identifiers on both iOS and Android have evolved.
Platforms provide app developers with identifiers with different properties, to be used for different purposes.
However, for the most part, these identifiers are designed to assist the developers/app providers rather than being user-facing. 
In the context of users exercising their legal rights, we argue that more needs to be done to raise awareness of issues relating to the rights of data subjects, as is a greater consideration of how technical mechanisms can assist users in exercising these rights.

\section{The user need for identifiers}



To recap, without an identifier which is both user-facing and collected by the app developer/provider, matching an individual to their data --- and therefore permitting them to exercise their rights --- may be difficult, and in some cases impossible. 
This is unless other methods have been considered and implemented into the app in advance, which relies on a prior awareness (and willingness) by those that designed the app.
Casual observation of various apps shows that exercising rights often requires the provision of further data, for instance, linking to a third-party sign-on service (e.g. Facebook), or forcing the registration of an account. 
This intuition is that this works to associate some identity information with the existing data profile, which then allows the individual to identify themselves (and therefore their data profile) to exercise their rights.

However, it is arguable that forcing an individual to provide more personal information --- or indeed, to sign up for an account just to exercise their rights --- runs against the spirit of data protection regulations. 
Given that such registration was otherwise unnecessary, this appears to serve more in the interests of the service provider than the user. 
That is, a provider may benefit from having another registered user, access to more data, and from service `lock-in' effects.  
In some ways, a forced registration process could be considered a `dark pattern'~\cite{fritsch2017}.
Further, it may act as a deterrent, as those who do not wish to provide such additional data may be discouraged from exercising their rights.

As such, there is a need for greater discussion and awareness about how users can best identify themselves in order to exercise their GDPR rights on mobile apps. 
We now outline two potential approaches to addressing this problem: one involving functionality built into the app itself, and another outlining how the platforms could implement features which facilitate the identification process.

\textbf{App providers should add functionality to help users identify themselves}

Clearly, there are steps which an app provider could take to facilitate the identification of users.
For example, identifiers could be exposed within the app itself.
This could be the actual identifier that the app provider uses internally, or perhaps an identifier created solely for the purposes of matching a user with their data.
The latter approach could allow for more user-friendly approaches, such as the use of mnemonic phrases (unique combinations of words), rather than pseudo-random alphanumeric strings of characters which users may have to carefully copy out. 
The identifier could be printed within the options of the app, such as in a \texttt{Help} or \texttt{Support} tab, or dynamically written into a privacy notice included within that app.

Such a feature may even make use of existing functionality, requiring limited or no further app development.
For example, we recently submitted a `right of access' (subject access) request to an app provider who then asked us to send a URL provided by the app's \texttt{Share} feature. 
This URL contained a unique code which was then used to identify our `nameless' profile from the app provider's system. 
By providing them with this URL, it enabled the identification of our data and our request was fulfilled.

Alternatively, and arguably better, would be to allow app providers to explore more proactive means for exercising rights. 
For example, providing functionality or a direct link within the app for exercising these rights; i.e. at the click of a button.
However, this requires each app provider to consider and implement these more involved features into their apps.

In practice, app providers must consider how they should respond to a rights request, and the technical measures they employ to facilitate this. 
However, at present, it appears that many app providers seem unaware of the issues, not least as some fairly simple methods could assist, for example, with user identification. 
To move forward, there is a need for greater levels of awareness regarding rights concerns, as well as a wider understanding about the incentives and advantages of thinking about technical measures for exercising such rights.
Regulatory pressure could also play a role as precedent begins to emerge around the GDPR's enforcement.

\textbf{Mobile platforms could allow users to view the identifiers used by their apps}

There is also scope for the platforms to facilitate data subjects exercising their rights, through simple changes to the way user identification is handled. 
Currently, there appears no practical way for a user to uncover iOS's IDFV and Android's app-specific SSAID for a given app on a device. 
From a platform perspective, functionality could be added that allows a user to view their allocated identifiers for all of the apps on their device. 
For instance, this could be presented to the user in a table from within the phone's settings menu, or under information regarding the installed app in question. 
This would ensure that any user could look up an app that they use and find the relevant identifier(s), which can assist controllers in identifying the relevant data for that given user. 
Retaining and including identifiers which are no longer current (i.e. uninstalled apps) within such a table could give the user full control over current and past profiles alike.
This simple functionality would go some way towards facilitating the exercising of rights, short of active intervention from the app providers themselves.

Of course, this is not a silver bullet; the identifier provided in this menu may not necessarily be an identifier that the app provider collects or uses. 
That said, it is common practice that app providers collect these platform-provided identifiers --- even if they are not the identifiers that are used internally by the app to build data profiles --- and this therefore represents a means to assist identification in many cases.

More involved steps could see the platforms requiring that apps use a particular identifier, or alternatively specify whichever identifier is being used (e.g. in an app's \texttt{manifest},\footnote{See: \textit{
App Manifest Overview} on the Android Developer Documentation~\cite{androiddocs2018}.}
or programatically).
These identifiers could then be made accessible to the user, as described above.
Given the platforms' application marketplaces entail a validation process (where access to other restricted identifiers is already checked), various methods could be employed to ensure conformance in a similar manner.

Regardless, having the mobile platforms taking even basic actions towards facilitating the exercising of data protection rights would raise the profile of data protection issues both with users and app providers alike. 
It would serve as a strong signal that providers need to take such concerns seriously. 
It may also begin a move towards a new and improved set of best practices for identification and rights management.
Moreover, building features which facilitate rights requests into the mobile platforms leaves little room for controllers to argue against non-compliance due to technical reasons or on grounds of excessive effort, particularly as regulators will quickly become aware of such functionality being added to the two main mobile platforms.

It remains to be seen whether large platforms would implement these features without a clear demand for such changes from either app providers/developers or end-users. 
Such demand may be slow to develop, given the apparent lack of awareness of these issues.
That said, even the basic solutions mentioned would be simple to implement and would go a long way towards empowering users and improving compliance with data protection regulations.

\section{Discussion}
There appears a need for improving the technical mechanisms that enable users to exercise their rights.
As the profile of data protection concerns continues to rise, best practices and design patterns are likely to emerge. 
This particularly given the GDPR's requirements for \emph{data protection by design and default}.\footnote{GDPR, Art 25(1).}

In terms of moving forward, having such features built directly into the platform gives some means for users to help identify themselves,\footnote{Again, it might not help in all cases, depending on whether the application actually records these specific platform-provided IDs.} regardless of whether the app provider explicitly provides mechanisms facilitating identity or data protection rights management.
Moreover, having such functionality explicitly built into the device's OS, accessible to users, highlights the importance of exercising rights, and raises users expectations regarding the apps they use. 
This works to incentivise conformance by app providers to either use or build-in means to help with identification and rights management. 
And by having such functionality available, where an app provider does not comply, users and regulators alike will know that mechanisms are possible, which works to limit excuses for non-compliance.

This discussion represents just the starting point.
There is a strong case for apps to have built-in means for directly facilitating the exercising of data protection rights, i.e. through the click of a button.
However, a
process for managing the validity of any rights requests is relevant for any app collecting personal data.
\textit{Is the user actually the individual? 
Has someone got hold of their unlocked phone?
}
For example, where there is no sign-in for an app, entering the device's lock\slash fingerprint\slash biometric screen could perhaps help. 
At the same time, while in-app authentication mechanisms may make sense in certain contexts, they could also act as a hindrance; such obstacles, depending on their implementation, might work to obstruct those wishing to exercise their rights.
Questions also arise in situations where a device is shared, such as a tablet used by several family members.

Another consideration is the degree of automation appropriate for exercising rights; for example, it may make sense to separate identity and rights functionality from the operation of the app, as a means to prevent someone masquerading as the user.
Also relevant are the security risks associated with identifiers: \textit{how should identifiers be managed, and what mitigations are in place if an identifier is leaked or otherwise compromised?}
All of the above issues require further thought, especially as an unwarranted rights request could cause far more harm than the general usage of the app. 
What is appropriate will of course depend on the circumstances and risk profile of the particular application.

Ultimately, exercising rights represents a legal obligation, and so there is both the need and opportunity for app providers and mobile platforms to do more.
Data subject identification is an ongoing discussion (see, for example, multi-user IoT devices~\cite{wachter2018} and web tracking through shadow profiles~\cite{garcia2017}), and there is much potential for future work in the space.
Raising awareness of the technical implications of rights is key, not least because technical mechanisms --- even those with privacy in mind --- may work to hinder individuals from exercising their rights~\cite{veale2018}.
By exploring such issues in a mobile app context, and suggesting simple actions that can be taken, our goal is to encourage new ways forward in data protection by design.

\section{Acknowledgements}
We acknowledge financial support from Microsoft Corporation for the Microsoft Cloud Computing Research Centre (MCCRC) and from the UK Engineering and Physical Sciences Research Council (EPSRC).

\balance{} 

\bibliographystyle{SIGCHI-Reference-Format}
\bibliography{reclaimingdata}

\end{document}